\documentclass[12pt]{article}
\title{\bf Hamiltonian approach to the bound state problem in $QCD_2$}
\author{
Yu.S.Kalashnikova\thanks{e-mail: yulia@vxitep.itep.ru}, A.V.Nefediev
\thanks{e-mail: nefediev@vxitep.itep.ru} and A.V.Volodin\thanks{e-mail: volodin@vxitep.itep.ru}}
\date{\it Institute of Theoretical and Experimental Physics,\\
117218, Moscow, Russia}
\newcommand{\be}{\begin{equation}}
\newcommand{\ee}{\end{equation}}

\newcommand{\too}{\mathop{\to}\limits_{N_C\to\infty}}
\newcommand{\vpint}{\int\makebox[0mm][r]{\bf --\hspace*{0.13cm}}}
\newcommand{\ds}{\displaystyle}
\newcommand{\vph}{\varphi}
\begin{document}
\maketitle

\begin{abstract}
Bosonization of the two-dimensional $QCD$ in the large $N_C$
limit is performed in the framework of Hamiltonian approach in the Coulomb 
gauge. The generalized Bogoliubov transformation is applied to diagonalize
the Hamiltonian in the bosonic sector of the theory, and the composite 
operators creating/annihilating bosons are obtained in terms of dressed quark
operators. The bound state equation is reconstructed as a result of the
generalized Bogoliubov transformation, and the form of its massless solution, 
chiral pion, is found explicitly. Chiral properties of the theory are discussed.
\end{abstract}

Two-dimensional quantum chromodynamics ($QCD_2$) in the large $N_C$
limit was first considered many years ago \cite{'tHooft} but
still remains popular 
in studies of various aspects of strong interactions. The reason for this
is two-fold. There are no transverse gluons in two dimensions, so the
theory is relatively simple; moreover, in the large $N_C$ limit only
planar graphs are to be summed, and a simple diagrammatic approach
can be developed. On the other hand, this theory does have nontrivial
content exhibiting both confinement and chiral symmetry breaking.

Most of the studies in $QCD_2$ were performed in the light-cone gauge,
which considerably simplifies the spectrum calculations but yields the
perturbative vacuum. 
Alternative approach based on the Coulomb gauge 
$A_1 = 0$ is more technically involved \cite{Bars&Green}, and it is not
a straightforward exercise to demonstrate the equivalence of the two
formulations. In particular, it appears that the vacuum is nontrivial 
in the Coulomb gauge version, and nonzero quark condensate exists 
for massless quarks \cite{Ming Li}. The latter feature is confirmed by the
sum rules approach \cite{Zhitnitsky} in the light-cone gauge. The confused
situation was resolved to large extent in the formulation on finite
intervals \cite{Lenz}, and the equivalence was demonstrated. 
 
In the present paper we study the Hamiltonian formulation of $QCD_2$ in 
the Coulomb gauge. In contrast to \cite{Lenz} the theory is quantized at
the ordinary time hypersurface. Our main purpose is to investigate some
peculiar properties of mesonic wave functions. It will be shown that 
the Bogoliubov--Valatin approach offers the most natural setting for
such studies. Finally we perform the generalized Bogoliubov transformation
and reformulate the theory in terms of the effective mesonic degrees of 
freedom. The presented approach allows straightforward calculations of 
any matrix elements of quark operators, and we demonstrate this property at
the example of evaluation of the pionic decay constant.

The Lagrangian of $QCD_2$ has the form
\be
L(x)= -\frac{1}{4}F^a_{\mu\nu}(x)F^a_{\mu\nu}(x)+\bar q(x)(i\hat{D}-m)q(x)
\ee
where $\hat{D} = (\partial_{\mu} - igA^a_{\mu}t^a)\gamma_{\mu}$, and
our convention for $\gamma$ matrices is $\gamma_0 = \sigma_3$,
$\gamma_1 = i\sigma_2$, $\gamma_5 = \gamma_0\gamma_1$. The large $N_C$
limit implies that $g^2N_C$ remains finite.

Gluonic propagator in the Coulomb gauge $A_1 = 0$ takes the form
$D_{00}(k_0,k) = - 1/k^2$, and the infrared singularity is regularized
by the principle value prescription yielding the linear confinement
\be
D^{ab}_{00}(x_0 - y_0, x - y ) =-\frac{i}{2}\delta^{ab} |x-y|\delta(x_0 - y_0).
\ee
  
Hamiltonian can be obtained by standard methods and reads 
\be
H=\int dxq^{+}(x)\left(-i\gamma^5\frac{\partial}{\partial x}+m\gamma^0\right) q(x)-
\frac{g^2}{2}\int dxdy\;q^{+}(x)t^aq(x)q^{+}(y)t^aq(y)\frac{\left|x-y\right|}{2},
\label{H}
\ee
where the quark and antiquark fields are defined by
\be
q_i(x_0,x)=\int\frac{dk}{2\pi}\left[u(k)b_i(x_0,k)+v(-k)d_i(x_0,-k)\right]e^{ikx}
\label{qi}
\ee
\be
b_i(k)|0\rangle= d_i(-k)|0\rangle =0
\label{bnd}
\ee
with
\be
u(k)=T(k)\left(1 \atop 0 \right)\quad v(k)=T(k)\left(0 \atop 1 \right)
\ee
$$
T(k)=e^{-\frac{1}{2}\theta(k)\gamma_1}.
$$

Parameter $\theta(k)$ has the meaning of Bogoliubov--Valatin angle
describing rotation from bare to dressed quarks.
Following \cite{Bars&Green} Hamiltonian (\ref{H}) can be normally
ordered in the basis of fermionic operators (\ref{bnd}):
\be
H=LN_C{\cal E}_v+:H_2:+:H_4:,
\ee
where ${\cal E}_v$ is the vacuum energy density ($L$ being the one-dimensional volume
of the space)
\be
{\cal E}_v =\int\frac{dp}{2\pi}Tr
\left\{\left(\gamma^5p+m\gamma^0\right)\Lambda_{-}(p)\right\}+\frac{\gamma}{4\pi}
\int\frac{dp\;dk}{(p-k)^2}Tr\left\{\Lambda_{+}(k)\Lambda_{-}(p)\right\}
\label{vac}
\ee
$$
\gamma=\frac{g^2}{4\pi}\left(N_C-\frac{1}{N_C}\right)\too \frac{g^2N_C}{4\pi}.
$$

Part $:H_2:$ is bilinear in quark fields:
$$
:H_2:=\int dx:q^{+}_i(x)\left(-i\gamma^5\frac{\partial}{\partial x}+m\gamma^0\right)q_i(x):
\hspace*{5cm}
$$
\be
\hspace*{3cm}-\frac{\gamma}{2}\int dxdy\frac{|x-y|}{2}\int dk :q^{+}_i(x)
\left[\Lambda_{+}(k)-\Lambda_{-}(k)\right]q_i(y):e^{ik(x-y)}
\ee
and the projection operators $\Lambda_{\pm}$ are introduced as
\be
\Lambda_{\pm} = T(k)\frac{1 \pm \gamma_0}{2}T^+(k).
\label{projectors}
\ee

The angle $\theta(k)$ is defined from the condition of diagonalizing
the bilinear in quark fields part of the Hamiltonian $:H_2:$. This condition
gives a system of integral equation for $\theta(p)$ and quark dispersion $E(p)$
\be
\left\{
\begin{array}{c}
E(p)cos\theta(p)=m+\frac{\ds \gamma}{\ds 2}\ds\vpint\frac{dk}{(p-k)^2}
cos\theta(k)\\
{}\\
E(p)sin\theta(p)=p+\frac{\ds \gamma}{\ds 2}\ds\vpint\frac{dk}{(p-k)^2}
sin\theta(k),
\end{array}
\right.
\label{system}
\ee
which leads to a gap equation for the angle $\theta(p)$ 
\be
p\cos\theta(p)-m\sin\theta(p)=\frac{\gamma}{2}\int\frac{dk}{(p-k)^2}\sin[\theta(p)-\theta(k)].
\label{gap}
\ee

Note that once the \lq\lq dressed" quarks (\ref{qi}), (\ref{bnd}) define excitations over
the true quark vacuum $|0\rangle$, then gap equation (\ref{gap}) could be 
reconstructed as the extremum condition 
\be
\frac{\delta {\cal E}_v[\theta]}{\delta\theta(p)}=0,
\ee
which ensures that the vacuum energy is minimal.

Thus the diagonalized Hamiltonian $:H_2:$ takes the form
\be
:H_2:=\int\frac{dk}{2\pi}
E(k)\left\{b_i^{+}(k)b_i(k)+d_i^{+}(-k)d_i(-k)\right\}.
\ee 

The obvious properties of all solutions of the system (\ref{system}) are 
$\theta(k) =
-\theta(-k)$, $E(k)= E(-k)$, and $\theta(k)\to\pi/2$ when $k\to\infty$.

Alternatively equation (\ref{gap}) could be found as the solution of the Schwinger--Dyson 
equations for the quark propagator:
\be
S(p_0,p)=\frac{1}{\hat{p}-m-\Sigma(p)}
\ee
\be
\Sigma(p)=-i\frac{\gamma}{2\pi}\int\frac{dk_0dk}{(p-k)^2}\gamma^0S(k_0,k)\gamma^0
=[E(p)\cos\theta (p)-m]+\gamma^1[E(p)\sin\theta (p)-p].
\ee

Gap equation (\ref{gap}) contains all information about one-fermion
sector of the theory, but not about the interaction between fermions,
because the $:H_4:$ part of the Hamiltonian 
was not involved so far. To proceed further we introduce,
as it was done in \cite{Lenz}, the colour singlet bilinear operators
\be
\begin{array}{c}
B(p,p')=\frac{\ds 1}{\ds\sqrt{N_C}}b_i^{+}(p)b_i(p')\quad 
D(p,p')=\frac{\ds 1}{\ds\sqrt{N_C}}d_i^{+}(-p)d_i(-p')\\
{}\\
M(p,p')=\frac{\ds 1}{\ds\sqrt{N_C}}d_i(-p)b_i(p')\quad
M^{+}(p,p')=\frac{\ds 1}{\ds\sqrt{N_C}}b^{+}_i(p')d^{+}_i(-p)
\end{array}
\ee
with commutation relations
$$
[M(p,p')M^{+}(q,q')]=-\frac{2\pi}{\sqrt{N_C}}\left\{
D(q,p)\delta(p'-q')+B(q',p')\delta(p-q)\right\}+
(2\pi)^2\delta(p'-q')\delta(p-q)
$$
\be
\too(2\pi)^2\delta(p'-q')\delta(p-q),
\label{mcom}
\ee
\be
\begin{array}{c}
[B(p,p')B(q,q')]=\frac{\ds 2\pi}{\ds\sqrt{N_C}}\left(B(p,q^{\prime})\delta(p'-q)-
B(q,p')\delta(p-q')\right)\too 0\\
{}\\

[D(p,p')D(q,q')]=\frac{\ds 2\pi}{\ds\sqrt{N_C}}\left(D(p,q^{\prime})\delta(p'-q)-
D(q,p')\delta(p-q')\right)\too 0\\
{}\\

[B(p,p')M(q,q')]=-\frac{\ds 2\pi}{\ds\sqrt{N_C}}M(q,p')\delta(p-q')\too 0\\
{}\\

[B(p,p')M^+(q,q')]=-\frac{\ds 2\pi}{\ds\sqrt{N_C}}M^+(p,q)\delta(p'-q')\too 0\\
{}\\

[D(p,p')M(q,q')]=-\frac{\ds 2\pi}{\ds\sqrt{N_C}}M(p',q')\delta(p-q)\too 0\\
{}\\

[D(p,p')M^+(q,q')]=-\frac{\ds 2\pi}{\ds\sqrt{N_C}}M^+(p,q')\delta(p'-q)\too 0.\\
\end{array}
\label{com}
\ee

In terms of these operators Hamiltonian (\ref{H}) can be presented as
$$
H=LN_C{\cal E}_v+\sqrt{N_C}\int\frac{dk}{2\pi}E(k)\{B(k,k)+D(k,k)\}
$$
$$
-\frac{\gamma}{2}\int\frac{dp\;dk\;dQ}{(2\pi)^2(p-k)^2}
\left[2\cos\frac{\theta(p)-\theta(k)}{2}\sin\frac{\theta(Q-p)-\theta(Q-k)}{2} 
\right.
$$
$$
\times\left\{M^{+}(p,p-Q)D(k-Q,k)+M^{+}(p-Q,p)B(k-Q,k)\right.\hspace*{3cm}
$$
$$
\hspace*{3cm}\left. -B(p,p-Q)M(k-Q,k)-D(p,p-Q)M(k,k-Q)\vphantom{M^+}\right\}
$$
\be
+\cos\frac{\theta(p)-\theta(k)}{2}\cos\frac{\theta(Q-p)-\theta(Q-k)}{2}
\label{HH}
\ee
$$
\times\left\{\vphantom{M^+}B(p-Q,p)B(k,k-Q)+D(p,p-Q)D(k-Q,k)\right.\hspace*{3cm}
$$
$$
\hspace*{3cm}+\left.M^{+}(p-Q,p)M(k-Q,k)+M^{+}(p,p-Q)M(k,k-Q)\right\}
$$
$$
+\sin\frac{\theta(p)-\theta(k)}{2}\sin\frac{\theta(Q-p)-\theta(Q-k)}{2}
$$
$$
\times\left\{\vphantom{M^+}B(p,p-Q)D(k,k-Q)+B(p-Q,p)D(k-Q,k)\right.\hspace*{3cm}
$$
$$
\left.\hspace*{3cm}\left.+M(p,p-Q)M(k-Q,k)+M^{+}(p-Q,p)M^{+}(k,k-Q)\right\}
\vphantom{\frac{dk}{(p-k)^2}}\right].
$$

It can be easily checked that the anzatz \cite{Lenz}
\be
\begin{array}{c}
B(p,p')=\frac{\ds 1}{\ds\sqrt{N_C}}\ds\int\frac{\ds dq''}{\ds 2\pi}M^{+}(q'',p)M(q'',p')\\
{}\\
D(p,p')=\frac{\ds 1}{\ds\sqrt{N_C}}\ds\int\frac{\ds dq''}{\ds 2\pi}M^{+}(p,q'')M(p',q'')
\end{array}
\label{anzatz}
\ee
satisfies the commutation relations (\ref{com}), so that in the leading in $N_C$ order
Hamiltonian (\ref{HH}) can be rewritten  as
$$
H=LN_C{\cal E}_v+\int\frac{dQdp}{(2\pi)^2}\left[(E(p)+E(Q-p))M^{+}(p-Q,p)M(p-Q,p)\right.
$$
\be
-\frac{\gamma}{2}\int\frac{dk}{(p-k)^2}\left\{2C(p,k,Q)M^{+}(p-Q,p)M(k-Q,k)\right.
\label{HHH}
\ee
$$
\left.\left.+S(p,k,Q)\left(M(p,p-Q)M(k-Q,k)+M^{+}(p,p-Q)M^{+}(k-Q,k)\right)\right\}\right],
$$
where
\be
\begin{array}{c}
C(p,k,Q)=\ds\cos\frac{\ds\theta(p)-\theta(k)}{\ds 2}\cos\frac{\ds
\theta(Q-p)-\theta(Q-k)}{\ds2}\nonumber\\
\vphantom{.}\\
S(p,k,Q)=\ds\sin\frac{\ds\theta(p)-\theta(k)}{\ds
2}\sin\frac{\ds\theta(Q-p)-\theta(Q-k)}{\ds2}\nonumber.
\label{cs}
\end{array}
\ee

We are now to perform the second Bogoliubov transformation which should
diagonalize this Hamiltonian. Let us define the new operators as
\be
\begin{array}{c}
m^{+}_n(Q)=\ds\int\frac{\ds dq}{\ds 2\pi}\left\{M^+(q-Q,q)\vph_+^n(q,Q)+
M(q,q-Q)\vph_-^n(q,Q)\right\}\\
\\
m_n(Q)=\ds\int\frac{\ds dq}{\ds 2\pi}\left\{M(q-Q,q)\vph_+^n(q,Q)+
M^+(q,q-Q)\vph_-^n(q,Q)\right\},
\label{m}
\end{array}
\ee
where functions $\vph^n_+$ and $\vph^n_-$ parametrize the given
transformation and obey Bogoliubov-type normalization and completeness
conditions $(m,~n>0)$:
\be
\begin{array}{rcl}
\ds\int\frac{\ds dp}{\ds 2\pi}\left(\vph_+^n(p,Q)\vph_+^{m}(p,Q)-\vph_-^n(p,Q)\vph_-^m(p,Q)
\right)&=&\delta_{nm}\\
&&\\
\ds\int\frac{dp}{2\pi}\left(\vph_+^n(p,Q)\vph_-^{m}(p,Q)-\vph_-^n(p,Q)\vph_+^m(p,Q)
\right)&=&0
\label{norms}
\end{array}
\ee
\be
\begin{array}{rcl}
\ds\sum\limits_{n=0}^{\infty}\left(\vph^n_+(p,Q)\vph^n_+(k,Q)-\vph^n_-(p,Q)
\vph^n_-(k,Q)\right)&=&2\pi\delta\left( p-k\right)\nonumber\\
&&\\
\ds\sum\limits_{n=0}^{\infty}\left(\vph^n_+(p,Q)\vph^n_-(k,Q)-\vph^n_-(p,Q)
\vph^n_+(k,Q)\right)&=&0\nonumber.
\label{complet}
\end{array}
\ee

It is easy to see using these conditions that in the $N_C\to\infty$ limit the new 
operators $m(Q)$ and $m^+(Q)$ obey standard bosonic commutation relations
\be
\left[m_n(Q)m^+_m(Q')\right]=2\pi\delta(Q-Q')\;\delta_{nm},\quad
\left[m_n(Q)m_m(Q')\right]=\left[m^+_n(Q)m^+_m(Q')\right]=0
\label{mcoms}
\ee
 
Straightforward but tedious calculation shows that the transformation (\ref{m}) 
diagonalizes Hamiltonian (\ref{HHH}) if functions $\vph^n_+$ and
$\vph^n_-$ are the solutions to the system of equations
\be
\left\{
\begin{array}{c}
[E(p)+E(Q-p)-Q_0]\vph_+(p,Q)\hspace*{8cm}\\
\hspace*{4.2cm}=\gamma\ds\vpint\frac{\ds dk}{\ds (p-k)^2}
\left[C(p,k,Q)\vph_+(k,Q)-S(p,k,Q)\vph_-(k,Q)\right]\\
{}\\

[E(p)+E(Q-p)+Q_0]\vph_-(p,Q)\hspace*{8cm}\\
\hspace*{4.2cm}=\gamma\ds\vpint\frac{\ds dk}{\ds (p-k)^2}
\left[C(p,k,Q)\vph_-(k,Q)-S(p,k,Q)\vph_+(k,Q)\right],
\end{array}
\right.
\label{BG}
\ee
and the resulting Hamiltonian takes the form
\be
H=LN_C{\cal E}'_v +\frac12\sum\limits_{n=0}^{+\infty}\int\frac{dQ}{2\pi}Q^0_n(Q)
\left\{m^+_n(Q)m_n(Q)+m_n(Q)m_n^+(Q)\right\},
\label{HHHH}
\ee
where $Q^0_n(Q)$ is the $n$-th positive eigenvalue of the system (\ref{BG}).
Hamiltonian (\ref{HHHH}) together with boson operators (20) comprises our procedure 
of bosonization. Note that here the vacuum energy density ${\cal E}'_v$ contains extra 
contribution which comes from the mesonic operators ordering which is of order $O(\frac{1}{N_C})$.

Equations (\ref{BG}) are nothing but those obtained by I.Bars and M.B.Green 
in \cite{Bars&Green} 
for the Bethe--Salpeter wave function
\be
\Phi(p,Q) = T(p)\left(\frac{1+\gamma_0}{2}\gamma_5\vph_+(p,Q) +
\frac{1-\gamma_0}{2}\gamma_5\vph_-(p,Q)\right)T^+(Q-p).
\label{phi}
\ee

There is a very important point concerning Bars--Green equations (\ref{BG}). 
While
equations (\ref{norms}), (\ref{complet}) are quite natural conditions 
imposed at the parameters
of Bogoliubov--Valatin transformations, it is apparently unacceptable
to assume such normalization and completeness conditions for the
solutions of bound state equations. In fact the problem is rooted in the 
properties of Bars--Green equations (\ref{BG}). It can be easily checked that
if system (\ref{BG}) is rewritten in the form of matrix integral 
Shr{\" o}dinger-like equation
\be
Q^n_0\left(\vph_{+}^n\atop \vph_{-}^n \right)=
\hat K\left(\vph_{+}^n\atop \vph_{-}^n \right)
\ee
then the integral operator $\hat K$ is not Hermitian. Fortunately
this does not cause a disaster as the eigenvalues are real: 
integrating both sides of (\ref{BG}) over $p$, doing
the same for complex conjugated equations and taking the appropriate linear
combination one arrives at
\be
\sum_{n=-\infty}^{+\infty}\left(Q_n^0-Q_m^{0*}\right)\int\frac{dp}{2\pi}
\left(\vph_+^n(p,Q)\vph_+^{m*}(p,Q)-\vph_-^n(p,Q)\vph_-^{m*}(p,Q)\right)=0,
\label{QQ}
\ee
that gives 
\be
Q^n_0 = Q^{n*}_0
\ee 
together with the orthonormality condition
\be
\begin{array}{rcl}
\ds\int\frac{\ds dp}{\ds 2\pi}\left(\vph_+^n(p,Q)\vph_+^{m}(p,Q)-\vph_-^n(p,Q)\vph_-^m(p,Q)
\right)&=&\delta_{nm}\\
&&\\
\ds\int\frac{dp}{2\pi}\left(\vph_+^n(p,Q)\vph_-^{m}(p,Q)-\vph_-^n(p,Q)\vph_+^m(p,Q)
\right)&=&0.
\label{norms_new}
\end{array}
\ee

Solutions of the system (\ref{BG}) come in pairs: for each eigenvalue
$Q^n_0$ with eigenfunction $(\vph^n_+,\vph^n_-)$ there exists another
eigenvalue $-Q^n_0$ with eigenfunction $(\vph^n_-,\vph^n_+)$. With
this symmetry equation (\ref{norms_new}) can be rewritten in the form (\ref{norms}) 
where only positive eigenvalues enter. Similarly in attempts to construct the
Green function for the system (\ref{BG}) the completeness (\ref{complet}) 
can be derived.

From the point of view of Bethe--Salpeter equation the reason for such an
unusual norm is the following. The matrix equation for the function 
$\Phi$ contains projectors (\ref{projectors}), so that $\Phi$ is subject to the
constraint
\be
\Lambda_+(p)\Phi(p,Q)\Lambda_+(Q-p)=\Lambda_-(p)\Phi(p,Q)\Lambda_-(Q-p) =0,
\label{ll}
\ee
as it is clear from (\ref{phi}). Thus both the norm and the completeness are defined
in the truncated space (\ref{ll}).

It is well known that Bogoliubov--Valatin transformation not only changes
the operators, the ground state is transformed too. Indeed the boson
annihilation operator $m_n(Q)$ does not annihilate the vacuum $|0\rangle$
defined by equation (\ref{bnd}). The explicit expression relating bosonic
($\left.\left|\Omega\right.\right\rangle$) and 
fermionic ($|0\rangle$) vacua is rather complicated, but luckily in the large $N_C$ 
limit the matrix elements of quark bilinears calculated in old and new
vacua coincide. For example, the chiral condensate
$$
\langle\bar{q}q\rangle=
\left\langle\Omega\left|\bar q_i(x)q^i(x)\right|\Omega\right\rangle\too
\left\langle 0\left|\bar q_i(x)q^i(x)\right|0\right\rangle\hspace*{5cm}
$$
\be
\hspace*{5cm}=N_C\int\frac{dk}{2\pi}Tr\left\{\gamma^0\Lambda_{-}(k)\right\}=
-\frac{N_c}{2\pi}\int\limits_{-\infty}^{+\infty}dk\cos\theta(k)
\label{condensate}
\ee
is the one found in \cite{Ming Li}.

Form (\ref{m}) for the operator $m^+_n(Q)$ suggests the obvious
particle--hole interpretation: the wave function of a meson moving
forward in time contains two pieces, of a quark--antiquark pair
moving forward in time with the amplitude $\vph_+$ and a pair moving
backward with the amplitude $\vph_-$. No effect of such a kind could 
ever happen in quark potential models.

What can be said about the relative size of the two amplitudes $\vph_+$ and
$\vph_-$? System (\ref{BG}) was solved numerically in \cite{Birse}, and 
it was shown that the $\vph_-$ component is small i) if the quark mass is large 
and ii) for higher excited states. In these two cases quark potential
model with local linear confinement serves as a good approximation. 

Besides $\vph_-$ component dies out with the increase of total mesonic 
momentum $Q$: in the
infinite momentum frame $(Q\to\infty)$ it is zero whereas the equation
for $\vph_+$ goes to the 't~Hooft equation \cite{'tHooft}
after appropriate rescaling, as it was shown in \cite{Bars&Green}. 

In conclusion, let us consider briefly the state which suffers from 
the effect of backward motion in the most dramatic way. It is two-dimensional
pion. It was shown in \cite{Ming Li} that the gap equation had a 
nontrivial solution in the chiral limit $m=0$, and the chiral condensate
(\ref{condensate}) did not vanish with this solution. So the Goldstone mode should
exist in the spectrum. Indeed, the set of functions
\be
\vph^{\pi}_{\pm}(p,Q) =N^{-1}_{\pi}\left(\cos\frac{\theta(Q-p)-\theta(p)}{2}\pm
\sin\frac{\theta(Q-p)+\theta(p)}{2}\right)
\label{pion}
\ee
is a solution of the system (\ref{BG}) for $Q_0=\sqrt{Q^2}$ and $Q>0$ 
(for $Q<0$ $\vph_{\pm} \Longleftrightarrow \vph_{\mp}$!). Here $N_{\pi}$ is the 
pion norm defined according to the general form (\ref{norms_new})
\be
N_{\pi}^2(Q)=\int_{-\infty}^{\infty}\frac{dp}{2\pi}\left[(\vph^{\pi}_+(p,Q))^2-
(\vph^{\pi}_-(p,Q))^2\right]
\ee
or with the solution (\ref{pion}) substituted one arrives at
\be
N_{\pi}^2(Q)=\int_{-\infty}^{\infty}\frac{dp}{\pi}[\sin\theta(p)+\sin\theta(Q-p)]
=\frac{2}{\pi}Q.
\label{pinorm}
\ee

For $Q=0$ $\vph_+(p,0) = \vph_-(p,0)\sim\cos\theta(p)$, pion spends half of time 
in the backward motion of the pair, and as follows from equation (\ref{pinorm})
such a function has zero norm, as it should be for the massless particle at rest.
In the opposite limiting case $Q\to\infty$ the backward motion part
dies out so that
\be
\vph^{\pi}_+(p,Q)\mathop{\to}\limits_{Q\to\infty}\sqrt{\frac{2\pi}{Q}},\quad 0\leq p\leq Q
\ee
coinciding with the Goldstone mode of the 't~Hooft equation. This does not
mean, nevertheless, that the pionic physics is exhausted by the simple
picture of linear confinement in the infinite momentum frame. All nontrivial
content of the wave function (\ref{pion}) is concentrated in the boundary
regions $x\to 0$ and $x\to 1$, $x = p/Q$. The same is true of
course for the $QCD_2$ quantized at the light-cone \cite{Zhitnitsky}, where
quantities like chiral condensate do not come out trivially,
and one is forced to employ the sum rules approach to arrive at a reliable
answer.  

With the given Hamiltonian approach it appears straightforward to calculate
any matrix elements of currents between mesonic states. For example,
to evaluate the pion decay constant $f_{\pi}$ defined in the standard way
\be
\left.\left\langle\Omega\right.\right|J_{\mu}^5(x)\left|\left.\pi(Q)\right.
\right\rangle=f_{\pi}Q_{\mu}\frac{e^{-iQx}}{\sqrt{2Q_0}},
\label{j5}
\ee
we express the axial-vector current $J_{\mu}^5(x)=\bar{\psi}(x)\gamma_{\mu}\gamma_5
\psi(x)$ in terms of mesonic creation--annihilation operators $m_n^+$, $m_n$ 
that allows to calculate the matrix element on the l.h.s. explicitly
and gives for $f_{\pi}$
\be
f_{\pi}=\sqrt{\frac{N_C}{\pi}}.
\ee

It is instructive to note that for any mesonic state ${\cal M}_n(Q)$ the analogous 
matrix element 
$\left.\left\langle\Omega\right.\right|J_{\mu}^5(x)\left|\left.{\cal M}_n(Q)\right.
\right\rangle$ contains this meson wave function integrated with the pionic one, that
obviously vanishes for any mesonic state but the pion when the pion norm (\ref{pinorm}) appears.
Thus the matrix element (\ref{j5}) is the only nontrivial one and the
corresponding decay constants for higher excited mesonic states vanish.

Now we can slightly relax the 
chiral limit and find the pion mass in the limit of small but not exactly
vanishing quark mass $m$. To this end we rewrite the system (\ref{BG}) 
as a single
equation for the matrix function $\Phi(p,Q)$ in the form
$$
Q_0\Phi(p,Q)=(\gamma_5p+\gamma_0m)\Phi(p,Q)-\Phi(p,Q)(\gamma_5(Q-p)+\gamma_0m)
\hspace*{5cm}
$$
\be
+\gamma\int\frac{dk}{(p-k)^2}\left\{\Lambda_+(k)\Phi(p,Q)\Lambda_-(Q-k)-
\Lambda_+(p)\Phi(k,Q)\Lambda_-(Q-p)\right.
\label{matrix}
\ee
$$
\hspace*{5cm}\left.-\Lambda_-(k)\Phi(p,Q)\Lambda_+(Q-k)+\Lambda_-(p)\Phi(k,Q)\Lambda_+(Q-p)\right\}.
$$

On multiplying equation (\ref{matrix}) by $\gamma_0\gamma_5$, taking trace 
over spinor indices and integrating over momentum $p$ one can arrive at 
\be
Q_0\int\frac{dp}{2\pi}Sp[\gamma_0\gamma_5\Phi(p,Q)]-
Q\int\frac{dp}{2\pi}Sp[\gamma_0\Phi(p,Q)]=
-2m\int\frac{dp}{2\pi}Sp[\gamma_5\Phi(p,Q)].
\ee

With the pion wave function (\ref{pion}) substituted and using definition (\ref{phi})
the latter equation gives a well-known relation \cite{Oakes} 
\be
f_{\pi}^2M_{\pi}^2=-2m\langle\bar{q}q\rangle,
\ee
so that one has for the pion mass
\be
M_{\pi}^2=2m\int_0^{\infty}dp\cos\theta(p)\sim m\sqrt{\gamma},
\ee
which vanishes in the exact chiral limit.

The last concluding remark concerning Hamiltonian (\ref{HHHH}) is in order.
As it could be anticipated from the very beginning this Hamiltonian describes
free non-interacting mesons, whereas the interaction suppressed by 
powers of $N_C$ is hidden in the terms present in (\ref{HH}) but 
omitted in (\ref{HHHH}). These terms define three and four meson vertices
so that their restoration gives quite a natural way of considering  
strong mesons decays and scattering amplitudes. This investigation is of a 
special interest in view of the fact that the pionic wave function 
is found explicitly and hence the \lq\lq mysterious" influence of the 
$q\bar q$ pair backward motion in time upon an exited meson decay into pions
can be easily resolved. This work is in progress and will be reported 
elsewhere in the future.
\medskip

We would like to thank A.A.Abrikosov Jr. for useful
discussions. Financial support of RFFI grants 97-02-16404 and 96-15-96740 is 
gratefully acknowledged.


\begin{thebibliography}{99}
\bibitem{'tHooft} G.'t~Hooft, Nucl.Phys. {\bf B75}, 461 (1974)
\bibitem{Bars&Green} I.Bars and M.B.Green, Phys.Rev. {\bf D17},
537 (1978)
\bibitem{Ming Li} Ming Li, Phys.Rev. {\bf D34}, 3888 (1986)
\bibitem{Zhitnitsky} A.R.Zhitnitsky, Sov.J.Nucl.Phys. {\bf 43},
999 (1986), {\bf 44}, 139 (1986)  
\bibitem{Lenz} F.Lenz and M.Thies, Ann.Phys. (N.Y.) {\bf 208}, 1
(1991)
\bibitem{Birse} Ming Li, L.Wilets and M.C.Birse, J.Phys. {\bf G13},
915 (1987)
\bibitem{Oakes} M.Gell-Mann, R.J.Oakes and B.Renner, Phys.Rev. {\bf 175} (1968) 
2195 
\end{thebibliography}
\end{document}